\documentclass[letter]{aa} 

\usepackage{diagbox}
\usepackage{xcolor}
\usepackage{soul}

\usepackage{float}
\usepackage{graphicx}
\usepackage{natbib}
\usepackage{txfonts}
\usepackage{braket}
\def \be {\begin{equation}}
\def \ee {\end{equation}}
\newcommand{\Zah}{\citetalias{zah77}}
\defcitealias{zah77}{Z77}

\newcommand{\Hut}{\citetalias{hut81}}
\defcitealias{hut81}{H81}

\newcommand{\Hur}{\citetalias{hur02}}
\defcitealias{hur02}{H02}

\begin{document}

\subtitle{Dynamical tides in binaries: Inconsistencies in the implementation of Zahn's prescription}

\author{Luca Sciarini\inst{1},
Sylvia Ekstr\"om\inst{1}, Patrick Eggenberger\inst{1}, Georges Meynet\inst{1,2}, Tassos Fragos\inst{1,2}, Han Feng Song\inst{3}}
\institute{Département d'Astronomie, Université de Genève, Chemin Pegasi 51, CH-1290 Versoix, Switzerland\\
              \email{luca.sciarini@unige.ch}
         \and
         Gravitational Wave Science Center (GWSC), Université de Genève, 24 quai E. Ansermet, CH-1211 Geneva, Switzerland
         \and
             College of Science, Guizhou University, Guiyang, 550025 Guizhou Province, PR China}

   \date{Received date 30 October 2023 /
Accepted date 6 December 2023}

 
  \abstract
{
  Binary evolution codes are essential tools to help in understanding the evolution of binary systems. They contain a great deal  of physics, for example  stellar evolution, stellar interactions, mass transfer, tides, orbital evolution. Since many  of these processes are difficult to account for in detail, we often rely on prescriptions obtained in earlier studies. We highlight that the impact of the dynamical tides with radiative damping has been implemented inconsistently with respect to its original theoretical formulation in many studies. We derive a new analytical solution for the evolution toward synchronization in the case of circular orbits and propose turnkey equations for the case of eccentric orbits that can be used in population synthesis studies. We compare the strength of the tidal torque obtained with this new formula with respect to that obtained with the formula generally used in literature by studying how the evolution toward synchronization of main sequence stellar models is affected. We conclude that by using an incorrect formula for the tidal torque, as has been done in many binary codes, the strength of the dynamical tides with radiative damping is over- or underestimated depending on whether the star is close to or far from synchronization. }
   
   


   \maketitle
  
%

\section{Introduction}
Tidal interaction is one of the crucial ingredients in binary system modeling. Despite the recent strides in tidal theory through the development of the resonance locking approach \citep[see, e.g.,][]{wit02,bur12,ful17,ma21,zan21}, the formalisms of \citet[][hereafter \Zah]{zah77} for dynamical tides and \citet[][hereafter \Hut]{hut81} for equilibrium tides remain largely used in detailed binary simulations and population synthesis studies. \Zah \
derived secular equations for tides in both radiative and convective regions, highlighting radiative damping on the dynamical tides and  viscous friction acting on the equilibrium tides as the primary dissipation mechanisms. 
However, the most widely used formalism for equilibrium tides is that of \Hut \ as their expressions are valid for any eccentricity. Even so,  \Hut\ did not treat dynamical tides with radiative damping, and therefore we still rely on \Zah\ in this case. In order to make the calculations of tides simpler to implement, there have been attempts to unify equilibrium and dynamical tides into a single framework.\par  

In \citet[][hereafter \Hur]{hur02}, the authors propose a derivation in which   dynamical and equilibrium tides are incorporated in a simple way into the \Hut\ formalism. This method has been extensively used in binary and triple modeling, and in population synthesis studies \citep[e.g.,][]{hur02, sep07, bel08, pax15, too16, qin18, fra23}.

In this letter we highlight that in many studies dynamical tides are implemented inconsistently with \Zah. We identify two types of inconsistencies. First, certain studies incorrectly implement the synchronization equation, deviating from the \Zah \ definition. Second, we point out that the \Hur\ derivation, although simple to implement, contains implicit approximations that make  it inconsistent with \Zah. In some cases this prescription can lead to results significantly different from those obtained with \Zah, which can impact many binary and triple  modeling results.
\section{Theory}\label{theory}
\subsection{Zahn and Hut formalisms}
\def\restriction#1#2{\mathchoice
              {\setbox1\hbox{${\displaystyle #1}_{\scriptstyle #2}$}
              \restrictionaux{#1}{#2}}
              {\setbox1\hbox{${\textstyle #1}_{\scriptstyle #2}$}
              \restrictionaux{#1}{#2}}
              {\setbox1\hbox{${\scriptstyle #1}_{\scriptscriptstyle #2}$}
              \restrictionaux{#1}{#2}}
              {\setbox1\hbox{${\scriptscriptstyle #1}_{\scriptscriptstyle #2}$}
              \restrictionaux{#1}{#2}}}
\def\restrictionaux#1#2{{#1\,\smash{\vrule height 1.\ht1 depth 1.\dp1}}_{\,#2}} 
The aim of this section is to emphasize the mathematical differences between the secular equations in the two cases,  equilibrium tides by \Hut\ and dynamical tides by \Zah,\ and to show why the two formalisms cannot be combined in a single description without losing consistency. We start by recalling the expressions obtained by \Hut\ in the case of equilibrium tides. They are given by equations (10) and (11) in their paper:
\begin{equation}
\scalebox{0.9}{$
\begin{split}
\restriction{\frac{\text{d}e}{\text{d}t}}{\rm Eq}&=-27\left(\frac{k}{T}\right)_{\rm c}q(1+q)\left(\frac{R}{a}\right)^8\frac{e}{(1-e^2)^{13/2}}\\&\qquad\cdot \left(f_3(e^2)-\frac{11}{18}(1-e^2)^{3/2}f_4(e^2)\frac{\Omega_{\rm spin}}{\Omega_{\rm orb}}\right),\\
\restriction{\frac{\text{d}}{\text{d}t}\left(I\Omega_{\rm spin}\right)}{\rm Eq}&=3\left(\frac{k}{T}\right)_{\rm c}MR^2q^2\left(\frac{R}{a}\right)^6\frac{\Omega_{\rm orb}}{(1-e^2)^{6}}\\&\qquad \cdot\left(f_2(e^2)-(1-e^2)^{3/2}f_5(e^2)\frac{\Omega_{\rm spin}}{\Omega_{\rm orb}}\right).
\end{split}
\label{hut}
$}
\end{equation}
Here $e$ is the eccentricity of the orbit, $\Omega_{\rm spin}$ and $\Omega_{\rm orb}$ the spin and orbital angular velocities, $k$ the apsidal motion constant, $T$ a typical timescale for orbital changes due to tidal evolution, $q=M_2/M$ the mass ratio, $I$ the moment of inertia, $R$ the stellar radius, $a$ the semimajor axis; $f_{2-5}$ are polynomials in $e^2$ whose definition are given in \Hut; the subscript $c$ refers to convective damping and the subscript Eq to equilibrium tides. From these equations, we can obtain the circularization and synchronization timescales defined as
\vspace{-0.3em}
\begin{equation}
\scalebox{0.9}{$
\begin{split}
\restriction{\frac{1}{\tau_{\rm circ}}}{\rm Eq}&\equiv-\restriction{\frac{\dot e}{e}}{e\approx 0,\ \Omega_{\rm spin}\approx\Omega_{\rm orb}}=\frac{21}{2}\left(\frac{k}{T}\right)_{\rm c}q(1+q)\left(\frac{R}{a}\right)^8,\\
\restriction{\frac{1}{\tau_{\rm sync}}}{\rm Eq}&\equiv-\restriction{\frac{\dot \Omega_{\rm spin}}{\Omega_{\rm spin}-\Omega_{\rm orb}}}{e\approx 0,\ I=\text{const}}=3\left(\frac{k}{T}\right)_{\rm c}\frac{MR^2}{I}q^2\left(\frac{R}{a}\right)^6.
\end{split}
\label{timescale}$}
\end{equation}
It should be noted that these timescales are obtained for $e\approx 0$ in both cases, $\Omega_{\rm spin}\approx\Omega_{\rm orb}$ in the case of the circularization timescale, and $I=$ const in the case of the synchronization timescale. In other words, we assume that the rotation of the star is already synchronized to the orbit and that the eccentricity is not too large to obtain the circularization timescale.\par We now recall the expressions obtained by \Zah\ for the time evolution of the same quantities in the case of the dynamical tides with radiative damping (i.e., equations (5.6) and (5.10) in their paper;\footnote{Our Eq. \eqref{zahn} slightly differs from Eqs. (5.6) and (5.10) in \Zah\, as the definition of the $s_{lm}$ is not used consistently throughout the whole paper, which is why it contains sign errors. We keep the original definition for  $s_{lm}$, given in equation (2.6) in \Zah, which is $s_{lm}=(l\Omega_{\rm orb}-m\Omega_{\rm spin})(R^3/GM)^{1/2}.$} they already assume $e\approx 0$),
\begin{equation}
\scalebox{0.9}{$
\begin{split}
\restriction{\frac{\text{d}e}{\text{d}t}}{\rm Dyn}&=-\frac{3}{4}e\left(\frac{GM}{R^3}\right)^{1/2}q(1+q)^{11/6}E_2\left(\frac{R}{a}\right)^{21/2}\\&\qquad\cdot \left(\frac{3}{2}-\frac{1}{4}(1+\zeta)^{8/3}-\zeta^{8/3}+\frac{49}{4}(1-\zeta)^{8/3}\right),\\
\restriction{\frac{\text{d}}{\text{d}t}\left(I\Omega_{\rm spin}\right)}{\rm Dyn}&=\frac{3}{2}\frac{GM^2}{R}E_2\left(q^2\left(\frac{R}{a}\right)^6\right)s_{22}^{8/3},
\end{split}
\label{zahn}$}
\end{equation}
where ``Dyn'' refers to dynamical tide, \begin{equation}
\scalebox{0.9}{$
\begin{split}
    s_{22}&=2(\Omega_{\rm orb}-\Omega_{\rm spin})(R^3/GM)^{1/2}, \\\zeta&=2(\Omega_{\rm spin}-\Omega_{\rm orb})/\Omega_{\rm orb}
    \end{split}$}.
\end{equation}
Using Eq. \eqref{zahn}, we obtain the circularization and synchronization timescale, defined as above:
\begin{equation}
\scalebox{0.9}{$
\begin{split}
\restriction{\frac{1}{\tau_{\rm circ}}}{\rm Dyn}&\equiv-\restriction{\frac{\dot e}{e}}{e\approx 0,\ \Omega_{\rm spin}\approx\Omega_{\rm orb}}\\&=\frac{21}{2}\left(\frac{GM}{R^3}\right)^{1/2}q(1+q)^{11/6}E_2\left(\frac{R}{a}\right)^{21/2},\\
\restriction{\frac{1}{\tau_{\text{ sync},s_{22}}}}{\rm Dyn}&\equiv-\restriction{\frac{\dot \Omega_{\rm spin}}{\Omega_{\rm spin}-\Omega_{\rm orb}}}{e\approx 0,\ I=\text{const}}\\&=3\left(\frac{GM}{R^3}\right)^{1/2}\frac{MR^2}{I}E_2\left(q^2\left(\frac{R}{a}\right)^6\right)s_{22}^{5/3}.
\end{split}
\label{s22}$}
\end{equation}
We indicate by the subscript $s_{22}$ in Eq. \eqref{s22} that the synchronization timescale is not constant, but depend on $s_{22}$ (i.e., on the difference $\Omega_{\rm spin}-\Omega_{\rm orb}$). This corresponds to the timescale $\tau_{\rm rot}$ in \Zah.\ In their paper, \Zah\ then introduce a new timescale, which they call $\tau_{\rm sync}$. The definition of $\tau_{\rm sync}$ for dynamical tides is not the same as in the case of equilibrium tides above in their paper, for which $\tau_{\rm sync}$ is defined as in Eq. \eqref{timescale}. We refer to this timescale as $\tau_{\rm sync,const}$. The idea of this timescale is that it is defined in a form that makes it independent of $s_{22}$. It is given by
\begin{equation}
\scalebox{0.9}{$
\begin{aligned}
\restriction{\frac{1}{\tau_{\text{sync,const}}}}{\text{Dyn}} & \equiv \restriction{\frac{\text{d}}{\text{d}t}\left|\frac{\Omega_{\text{spin}}-\Omega_{\text{orb}}}{\Omega_{\text{orb}}}\right|^{-5/3}}{e\approx 0,\ \Omega_{\text{orb}}=\text{const},\ I=\text{const}} \\
&= 5\cdot 2^{5/3}\left(\frac{GM}{R^3}\right)^{1/2}\frac{MR^2}{I}q^2(1+q)^{5/6}E_2\left(\frac{R}{a}\right)^{17/2}.
\end{aligned}
$}
\label{const}
\end{equation}
This definition for the synchronization timescale is used only in the case of the dynamical tide, as it is the only case where the difference in $\Omega$ does not appear linearly in the angular velocity equation. Looking at the equations used in some papers that use the \Zah\ prescription \citep[e.g.,][]{cla04, sie13}, it seems there has sometimes been a confusion between $\tau_{\rm sync,const}$ and $\tau_{{\rm sync},s_{22}}$. These authors  used $1/\tau_{{\rm sync}}=-\frac{\dot \Omega_{\rm spin}}{\Omega_{\rm spin}-\Omega_{\rm orb}}$ as the definition for the synchronization timescale, but they assigned the expression of $\tau_{\rm sync,const}$ (Eq. (7) in \citealt{cla04} and Eq. (29) in \citealt{sie13}). This confusion may have arisen from the fact that in the original paper (\Zah), the definition of $\tau_{\rm sync}$ is not kept consistent for the two cases (dynamical and equilibrium tides).
\subsection{Hurley 2002 derivation}
\citet{hur02} proposed a derivation that provides a ratio $\left(\frac{k}{T}\right)_{\rm rad}$ in the radiative case, which is incorporated into the  \Hut\ model of equilibrium tides (see Eq. \eqref{hut}). It relies on the comparison of the circularization timescales given in Eqs. \eqref{timescale} and \eqref{s22} in order to obtain the following expression for the ratio
\begin{equation}
 \left(\frac{k}{T}\right)_{\rm rad}=\sqrt{\frac{GMR^2}{a^5}}(1+q)^{5/6}E_2
\label{ratio}
\end{equation} in the radiative case. Inserting this expression into Eq. \eqref{timescale} in the radiative case gives an expression for the synchronization timescale, which is very similar (up to a constant factor) to that in Eq. \eqref{const}: 
\begin{equation}
\scalebox{0.9}{$
\begin{split}
\restriction{\frac{1}{\tau_{\text{ sync,hur}}}}{\rm Dyn}&\equiv-\restriction{\frac{\dot \Omega_{\rm spin}}{\Omega_{\rm spin}-\Omega_{\rm orb}}}{e\approx 0,\ I=\text{const}}\\&=3\left(\frac{GM}{R^3}\right)^{1/2}\frac{MR^2}{I}q^2(1+q)^{5/6}E_2\left(\frac{R}{a}\right)^{17/2}.
\end{split}
\label{falsetime}$}
\end{equation}
We refer to this timescale as $\tau_{\rm sync,hur}$, although in their paper \Hur\ do not obtain the same factor in front of the expression, but instead have the $5\cdot 2^{5/3}$ factor that comes from \Zah.\ Comparing Eqs. \eqref{const} and \eqref{falsetime}, we note that the expressions of $\tau_{\rm sync,const}$ and $\tau_{{\rm sync,hur}}$ are almost the same, although they have different mathematical definitions. On the other hand, $\tau_{{\rm sync},s_{22}}$ and $\tau_{{\rm sync,hur}}$ have different mathematical expressions, even though their definitions seem to be the same. This is due to the different ways they are obtained. In Appendix \ref{AppA} we show a comparison of $\tau_{{\rm sync},s_{22}}$ and $\tau_{{\rm sync,hur}}$ as a function of the difference in $\Omega$.\par The problem in the derivation of Eq. \eqref{falsetime} is that it relies on the circularization timescales, which are calculated under the simplification that synchronization has already been achieved:  $\Omega_{\rm spin}\approx\Omega_{\rm orb}$. Under this assumption the respective dependencies in $\Omega_{\rm spin}$ and $\Omega_{\rm orb}$ (which are very different) in Eq. \eqref{hut} and \eqref{zahn} drop, which allows a somewhat straightforward match between the two circularization timescales. This approach is inconsistent, as in a sense synchronization is assumed in order to obtain an equation for the evolution toward synchronization. One may wonder whether this can be just seen as an additional approximation, and that in the limit $\Omega_{\rm spin}\approx \Omega_{\rm orb}$, the evolution obtained following \Hur\ should converge to the \Zah\ prescription. But one can actually see that since the difference in $\Omega$ is raised to the power 5/3 in Eq. \eqref{s22}, there is no Taylor expansion around $\Omega_{\rm spin}\approx \Omega_{\rm orb}$, and  thus the equation does not converge to Eq. \eqref{falsetime} when $\Omega_{\rm spin}\approx \Omega_{\rm orb}$. The point is that the approximation $\Omega_{\rm spin}\approx \Omega_{\rm orb}$ is applied in the \Hur\ derivation to the eccentricity equation, not the angular velocity equation. \par Finally, we note that the  \Hut\ formalism was derived for equilibrium tides. Even though comparing the two circularization timescales in order to obtain a value for $(k/T)_{\rm rad}$, as done by \Hur, seems reasonable, this method would not work if the circularization timescales were not approximated beforehand, as the respective dependencies in $\Omega$ would not drop. As these differences are precisely present because of the different nature of the tides, it does not seem     justified to make them disappear.\par For these reasons it makes much more sense in our view to directly use \Zah\ (i.e., Eq. \eqref{zahn}) in order to compute the time evolution of $\Omega_{\rm spin}$ and $e$ in the case of dynamical tides with radiative damping.
\subsection{Extension and adjustments of \citet{zah77}}Equation \eqref{zahn} has a few flaws, which  are discussed in Appendix \ref{AppB}. In the rest of the letter we use the following secular equations adapted from the  \Zah\ paper, whose derivations are given in Appendix \ref{AppB}. These are the equations we would recommend to use for the dynamical tides when the eccentricity is nonnegligible:
\vspace{-0.5em}
\begin{equation}
\scalebox{0.9}{$\begin{split}
&\restriction{\frac{\text{d}}{\text{d}t}\left(I\Omega_{\rm spin}\right)}{\rm Dyn}=\frac{3}{2}\frac{GM^2}{R}E_2\left(q^2\left(\frac{R}{a}\right)^6\right)\cdot\Bigg\{s_{22}^{8/3}\text{sgn}(s_{22})\\&+e^2\left(\frac{1}{4}s_{12}^{8/3}\text{sgn}(s_{12})-5s_{22}^{8/3}\text{sgn}(s_{22})+\frac{49}{4}s_{32}^{8/3}\text{sgn}(s_{32})\right)\Bigg\}\\
&\restriction{\frac{\text{d}e}{\text{d}t}}{\rm Dyn}=-\frac{3}{4}e\left(\frac{GM}{R^3}\right)^{1/2}q(1+q)^{1/2}E_2\left(\frac{R}{a}\right)^{13/2}\cdot \\&\Bigg(\frac{3}{2}s_{10}^{8/3}\text{\rm sgn}(s_{10})-\frac{1}{4}s_{12}^{8/3}\text{\rm sgn}(s_{12})-s_{22}^{8/3}\text{\rm sgn}(s_{22})+\frac{49}{4}s_{32}^{8/3}\text{\rm sgn}(s_{32})\Bigg).
\end{split}
\label{zahn_corrected}$}
\end{equation}
\subsection{Analytical solutions}\label{analytical}
Analytical solutions can be obtained in the case of circular orbits (i.e., for both Eqs. \eqref{const} and \eqref{falsetime}). We note that the analytical solution of Eq. \eqref{const} is not found in the  literature and should be used instead of that of Eq. \eqref{falsetime} in detailed binary simulations and population synthesis studies. We obtain them using the same expression for the synchronization timescale $1/\tau_{\rm sync}=5\cdot 2^{5/3}\left(\frac{GM}{R^3}\right)^{1/2}\frac{MR^2}{I}q^2(1+q)^{5/6}E_2\left(\frac{R}{a}\right)^{17/2}$, which is the expression generally used in the literature (the expression of $1/\tau_{\rm sync,hur}$ differs from this one by a constant factor).\par
\begin{figure}[h]
\centering
\centerline{\includegraphics[trim=0cm 0cm 2cm 1.5cm, clip=true, width=1.05\columnwidth,angle=0]{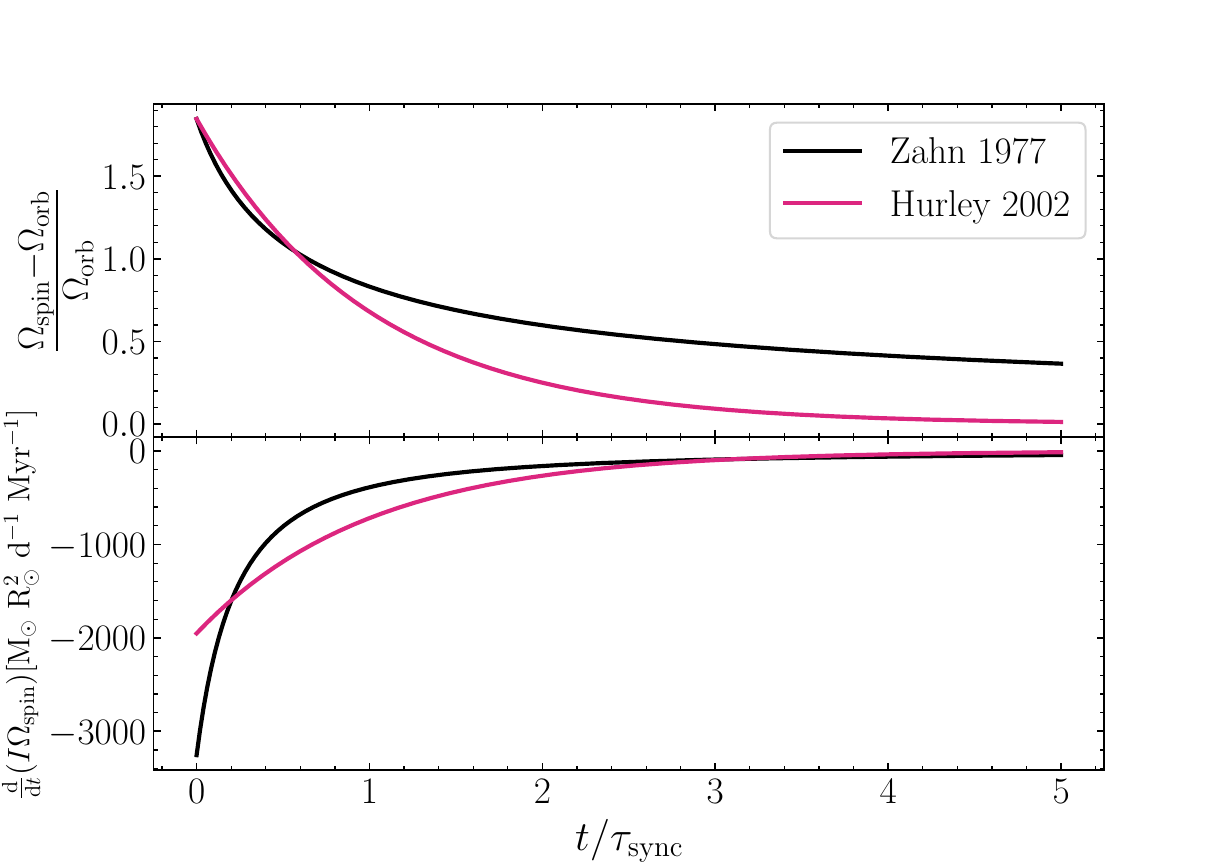}}
\caption{Evolution toward synchronization depending on the adopted definition for $\tau_{\rm sync}$.
\textit{Upper panel:} Analytical solutions of Eqs. \eqref{const} (blue curve) and \eqref{falsetime} (orange curve) as functions of $t/\tau_{\rm sync}$.\ \ \textit{Lower panel:} Torques as functions of $t/\tau_{\rm sync}$.}
\label{an_sol}
\end{figure}

 When the definition in Eq. \eqref{const} is used, the analytical solution is a power law
\vspace{-0.5em}
\begin{equation}
\frac{\Omega_{\rm spin}(t)-\Omega_{\rm orb}}{\Omega_{\rm orb}}=\left(\left(\frac{\Omega_{0}-\Omega_{\rm orb}}{\Omega_{\rm orb}}\right)^{-5/3}+\text{sgn} (\Omega_0-\Omega_{\rm orb})\frac{t}{\tau_{\rm sync}}\right)^{-3/5},
\end{equation}
where $\Omega_0=\Omega_{\rm spin}(t=0)$ and assuming that $\Omega_{\rm orb}$ and $\tau_{\rm sync}$ are constant.  When the definition in Eq. \eqref{falsetime} is used, the analytical solution is an exponential decay:
\begin{equation}
\frac{\Omega_{\rm spin}(t)-\Omega_{\rm orb}}{\Omega_{\rm orb}}=\left(\frac{\Omega_{0}-\Omega_{\rm orb}}{\Omega_{\rm orb}}\right)\exp\left(-\frac{t}{\tau_{\rm sync}}\right).
\end{equation}
In Fig. \ref{an_sol} we compare the analytical solutions for typical values for a system of two $M=15$ M$_{\odot}$ stars at zero age main sequence. We use $R=3.3\cdot 10^{9}$ m, $r_g^2=I/MR^2=0.075$, $E_2=3.4\cdot 10^{-6}$, and $P=2.3$ days, giving $\Omega_{\rm orb}=\frac{2\pi}{P}=3.16\cdot 10^{-5}$ rad/s, $a=\left(\frac{G(M+M_2)}{\Omega_{\rm orb}^2}\right)^{1/3}\approx 0.11$ au, and $\tau_{\rm sync}\approx 6.5\cdot 10^4$ years. We give for the initial condition $\Omega_0 =9\cdot 10^{-5 }$ rad/s. In the upper panel the relative departure from synchronism,  $\frac{\Omega_{\rm spin}-\Omega_{\rm orb}}{\Omega_{\rm orb}}$ is depicted as a function of time. In the lower panel, the corresponding torques $\frac{\text{d}}{\text{d}t}(I\Omega_{\rm spin})$ can be seen. The Zahn torque was computed consistently with Eq. \eqref{zahn_corrected} in the case $e\approx 0$, 
\begin{equation}
\frac{\text{d}}{\text{d}t}\left(I\Omega_{\rm spin}\right)_{\rm zahn}=\frac{3}{2}\frac{GM^2}{R}E_2\left(q^2\left(\frac{R}{a}\right)^6\right)\text{sgn}(s_{22})s_{22}^{8/3}
\label{zahn_torque}
,\end{equation}
and the Hurley torque was computed consistently with Eq. \eqref{falsetime}, but with the expression of the timescale of \eqref{const}:
\begin{equation}
\begin{split}
\frac{\text{d}}{\text{d}t}\left(I\Omega_{\rm spin}\right)_{\rm hur}&=5\cdot 2^{5/3}\left(\frac{GM}{R^3}\right)^{1/2}MR^2q^2(1+q)^{5/6}E_2\left(\frac{R}{a}\right)^{17/2}\\&\cdot(\Omega_{\rm orb}-\Omega_{\rm spin}).
\end{split}
\label{hurley_torque}
\end{equation}
We see by comparing the two curves in Fig. \ref{an_sol} that depending on the adopted definition for the synchronization timescale the evolution toward synchronization is very different, which is not surprising as the two analytical solutions have different mathematical expressions. Using the definition of \Hur\, (i.e., Eq. \eqref{falsetime}), the solution is an exponential decay, which decreases much faster. This implies that a model  of dynamical tides using the definition of $\tau_{\rm sync}$ as in Eq. \eqref{falsetime} would in general lead to an overestimation of the strength of the tides. More precisely, the  Zahn expression for the torque has a stronger dependence on the difference in $\Omega$ than the Hurley expression. When the star is far from synchronism ($t/\tau_{\rm sync}\lesssim 0.2$), this results in  a stronger torque with the Zahn expression. However, as the star gets close to synchronization, the small difference in $\Omega$ being raised to the power 8/3 induces a much smaller torque, as   can be seen in Fig. \ref{an_sol} when $t/\tau_{\rm sync}\gtrsim 0.2$.
\section{Stellar models}\label{models}
In this section we show the results of main sequence (MS) stellar evolution using the two different expressions for the tidal torque:  Eqs. \eqref{zahn_torque} and \eqref{hurley_torque}. We restrict ourselves to the case of circular orbits, but intend to extend this work to eccentric systems in a follow-up study.
\subsection{Ingredients of the stellar models}
We used the GENeva Evolutionary Code (GENEC; \citealt{egg08}) to compute the stellar evolution from the onset of central H burning and stopped the simulation when the star filled its Roche lobe. In order to obtain a strong core-envelope coupling (i.e., a case close to solid body rotation), the transport of angular momentum (AM) was computed accounting for the effects of the Tayler–Spruit dynamo (\citealt{spr02}, \citealt{mae04}). Apart from the AM transport, the stellar physics was handled as in \citet{eks12}. The size of the convective core was increased
with a step overshoot scheme with $\alpha_{\rm ov}=0.1$.\par We modeled the impact of dynamical tides as in \citet{son13,son16,son18}, who implemented the tides in GENEC. For each model, we computed a comparative model with the Hurley torque.\par In order to obtain a case close to the theoretical one, where $M$, $R$, and the other quantities in the right hand side of Eq. \eqref{const} remain constant throughout the evolution, we did not include stellar winds in the  simulations. We kept the orbital velocity constant throughout the simulation, which in this case is equivalent to assuming that the separation does not evolve. For $E_2$, we used the latest prescription from \citet{qin18} for MS stars:
\begin{equation}
E_2 = 10^{-0.42}\cdot (R_{\rm conv}/R)^{7.5}
.\end{equation}
\subsection{Spin-down case}
\begin{figure}[h]
\centering
\centerline{\includegraphics[trim=2cm 1cm 2cm 1.5cm, clip=true, width=1.02\columnwidth,angle=0]{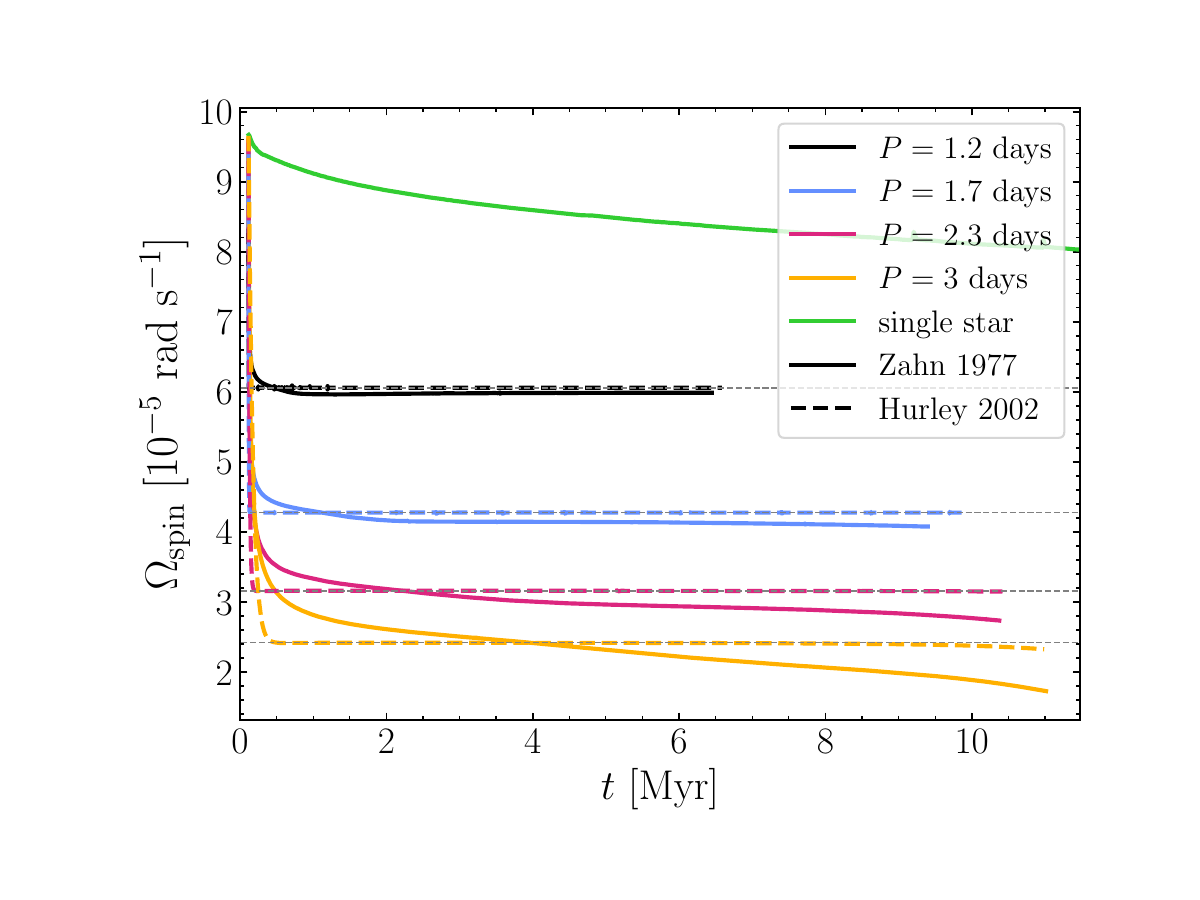}}
\caption{
MS $\Omega_{\rm spin}$ evolution in the spin-down case for stellar models with initial velocity $v/v_{\rm crit}=0.5$ and orbital periods $P=[1.2,1.7,2.3,3]$ days with the tidal torques defined in Eq. \eqref{zahn_torque} (Zahn 1977, solid line) and \eqref{hurley_torque} (Hurley 2002, dashed line). The grey dashed lines represent the orbital angular velocities corresponding to the selected periods. The single star evolution is shown as a reference.}
\label{spin_down}
\end{figure}

The time evolution of the surface angular velocity ($\Omega_{\rm spin}$) of 15 M$_{\odot}$, $Z=$ Z$_\odot$ stellar models with a companion of equal mass, circular orbits, and orbital periods $P=[1.2,1.7,2.3,3]$ days, corresponding to orbital angular velocities $\Omega_{\rm orb}=[6.06,4.28,3.16,2.42]\cdot 10^{-5}$ rad/s is presented in Fig. \ref{spin_down}. We give the stellar models an initial velocity of $v_0/v_{\rm crit}=0.5$ (where $v_{\rm crit}$ is the critical velocity computed as in \citealt{eks12}), which corresponds to an angular velocity of $\Omega_0=9.6\cdot 10^{-5}$ rad/s (i.e., the spin-down case). Spin-up cases are discussed in Appendix \ref{AppC}. We compare the evolution of the surface angular velocity depending on the  formula used for the torque, and we show as a reference the evolution of the surface angular velocity for the single star evolution. \par The solid lines represent the evolution obtained with the Zahn torque (Eq. \eqref{zahn_torque}) and the dashed lines the evolution with the Hurley torque (Eq. \eqref{hurley_torque}).
We note the following differences comparing the evolution obtained with the two prescriptions. When the original \Zah \ formula is used, the models reach synchronization later and do not remain synchronized for very long, in particular for longer periods, where the tides are less strong. For all periods except $P=1.2$ days, the surface angular velocity starts deviating from the orbital angular velocity, as the tides are not strong enough to keep it synchronized. This can be explained by the fact that when the stellar evolution is taken into account, the angular velocity is also modified by the evolution: the natural evolution of $\Omega_{\rm spin}$ during the MS is a decrease due to the expansion of the radius and the conservation of AM. This natural decrease in $\Omega_{\rm spin}$ can be observed looking at the single star evolution (cyan line). When tides are taken into account, the evolution toward (or away from) synchronization depends on the strength of the two processes affecting $\Omega_{\rm spin}$:  the decrease due to the expansion, and the tidal torque which tends to synchronize the angular velocity to the orbital one. As already noted in section \ref{analytical}, one of the main differences between the two torques is that in the Zahn torque in   Eq. \eqref{zahn}, the difference in $\Omega$ is raised to the power 8/3, whereas it only appears linearly in the Hurley torque \eqref{hurley_torque}. When this difference becomes small (i.e., when the star is close to synchronization), the power 8/3 implies that the Zahn torque becomes much smaller than the Hurley torque. The torque can be insufficient to keep the star synchronized, which can be seen in the late evolution of the stars with $P>1.2$ days. The fact that the difference in $\Omega$ only appears linearly in the Hurley torque implies that this expression is more efficient at maintaining the star close to synchronization. \par
\begin{figure}[h]
\centering
\centerline{\includegraphics[trim=1.5cm 1cm 2cm 1.5cm, clip=true, width=1.02\columnwidth,angle=0]{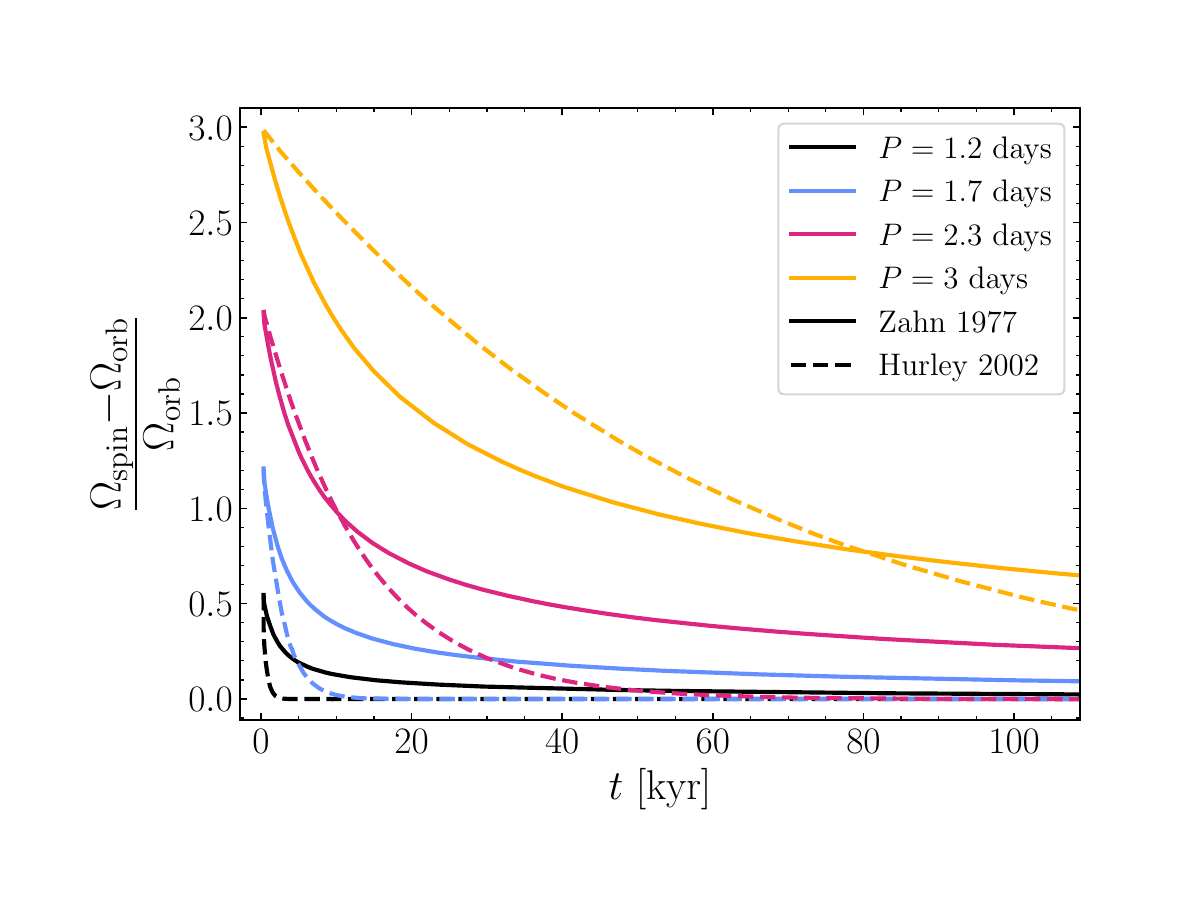}}
\caption{
Evolution for $t\lesssim 100$ kyr of the relative departure from synchronization $\frac{\Omega_{\rm spin}-\Omega_{\rm orb}}{\Omega_{\rm orb}}$ in the spin-down case for stellar models with initial velocity $v/v_{\rm crit}=0.5$ and orbital periods $P=[1.2,1.7,2.3,3]$ days with the tidal torques defined in Eqs. \eqref{zahn_torque} (Zahn 1977, solid line) and \eqref{hurley_torque} (Hurley 2002, dashed line).}
\label{relative_spin_down}
\end{figure}

 In Fig. \ref{relative_spin_down} we study in greater detail the evolution of the relative departure from synchronism (defined as in section \ref{analytical}) at the beginning of the simulations.
Consistently with the analytical solutions derived in section \ref{analytical}, we observe that in the $P=2.3$ days case, the Zahn expression for the torque implies a stronger decay of $\frac{\Omega_{\rm spin}-\Omega_{\rm orb}}{\Omega_{\rm orb}}$ at the beginning of the evolution, as the difference in $\Omega$ is initially large. When the difference becomes smaller, the situation is inverted and the Hurley torque becomes stronger. This is even more visible for the $P=3$ days case, where the difference in $\Omega$ is initially greater (as $\Omega_{\rm orb}$ is smaller). In the $P=1.2,1.7$ days cases, we do not observe any regime in which the Zahn torque is stronger ($\Omega_{\rm spin}$ decreases faster with the Hurley torque from the start). This is due to the fact that as the initial periods are smaller, the  $\Omega_{\rm orb}$ values are larger, and thus closer to $\Omega_0$, or equivalently the differences in $\Omega$ are small from the start. In any case, we see that the synchronization is systematically reached more quickly with the Hurley formula.
\section{Conclusions and discussions}\label{conclusion}
The present results show that the adopted definition for the synchronization timescale, or equivalently the adopted equation for the tidal torque in the case of the dynamical tides in radiative zones has an impact on the evolution toward synchronization. This is not surprising given the analytical solutions provided in the simple case of circular orbits. As an inconsistent formula has  been widely used in the literature, the strength of the dynamical tides has been misestimated in many studies.\par 
Overall, the conclusion of our analysis is that whenever the  \Hur\ prescription is used (or equivalently if the definition of the synchronization timescale is taken inconsistently with equation (5.8) in \Zah), the strength of the tidal torque is underestimated when the star is far from synchronism, and overestimated when it gets close to synchronism compared to the original \Zah\ prescription.
\par The \Hur\ prescription for the computation of the parameter $E_2$ was already challenged in earlier studies \citep[e.g.,][]{sie13,qin18,mir23}, which find that the \Hur\ prescription of $E_2$ in general overestimates the strength of the dynamical tides as it does not decrease during the main sequence. The present letter goes further by showing that under their formalism the synchronization timescale is inconsistent with \Zah .
\par This can have important consequences in the modeling of the evolution of  close binaries. Here are a few qualitative predictions that can be made based on the results of this study. They will be explored in greater detail in future works.\par We expect that when using the original \Zah \ equation for the torque the tides will be more efficient in spinning down the star when it is far from synchronism, which is typically the case during a mass transfer episode where the secondary quickly reaches critical velocity because of the accretion of AM. This is known as the angular momentum problem \citep{pac81}. We believe that by using the original formula for the tidal torque, the secondary would lose more AM than  is obtained with the Hurley torque. As the secondary moves far away from synchronism, the regime where the Zahn torque becomes stronger is reached. By removing more AM by tidal interaction, the efficiency of the mass transfer should increase.\par 

Another area where the revision presented here could have implications is in the question of the spin of the second-born black hole (BH) in the merging of binary black holes (BBHs). Several groups \citep[e.g.,][]{kus16, hot17, qin18, zal18, bav20,bel20} have studied the impact of the tides on the spin of the second-born black hole in stripped helium star plus black hole systems. This question is of particular interest since the effective inspiral spin parameter $\chi_{\rm eff}$ is reasonably well constrained in gravitational wave detection \citep{abb19}. As the spin of the first-born BH is expected to be low \citep[e.g.,][]{fra15,bel20,bav23}, $\chi_{\rm eff}$ is believed to be proportional to the spin of the second-born BH. In this case,  there is an anti-correlation between the merging timescale of the BBHs, $T_{\rm merger}$, and the spin of the second born BH   \citep[e.g.,][]{bav20}. This can be understood by the fact that in order for the merging timescale to be short, small separations are needed, which increases the strength of the tides and tends to increase the spin of the helium star  as the corresponding orbital angular velocities are high. 
Since several of these studies \citep{qin18, bav20,bel20,bav23}  used \Hur\ formulation for dynamical tides, we can expect that the fraction of tidally spun-up BH progenitors and the exact shape of predicted distribution of $\chi_{\rm eff}$ will be affected by the inconsistencies in the derivation highlighted in this letter. Quantifying the effect requires  carrying out binary population synthesis calculations. 
We leave the more in-depth exploration of this question for a follow-up study.
\begin{acknowledgements}
LS and SE have received support from the SNF project No 212143. SE, PE and GM have received funding from the European Research Council (ERC) under the European Union's Horizon 2020 research and innovation program (grand agreement No 833925, project STAREX). TF acknowledges support for this work through the Swiss National Science Foundation (project numbers PP00P2\_211006 and CRSII5\_213497). HS has received support from the National Natural Science Foundation of China (Grant Nos. 11863003, 12173010).  
\end{acknowledgements}

%
%

\bibliographystyle{aa}
\bibliography{myrefs}\begin{appendix}

\begin{thebibliography}{34}
\expandafter\ifx\csname natexlab\endcsname\relax\def\natexlab#1{#1}\fi

\bibitem[{Abbott {et~al.}(2019)Abbott, Abbott, Abbott, Abraham, Acernese,
  Ackley, Adams, Adhikari, Adya, Affeldt, Agathos, Agatsuma, Aggarwal, Aguiar,
  Aiello, Ain, Ajith, Allen, Allocca, Aloy, Altin, Amato, Ananyeva, Anderson,
  Anderson, Angelova, Antier, Appert, Arai, Araya, Areeda, Ar\`ene, Arnaud,
  Arun, Ascenzi, Ashton, Aston, Astone, Aubin, Aufmuth, AultONeal, Austin,
  Avendano, Avila-Alvarez, Babak, Bacon, Badaracco, Bader, Bae, Baker,
  Baldaccini, Ballardin, Ballmer, Banagiri, Barayoga, Barclay, Barish, Barker,
  Barkett, Barnum, Barone, Barr, Barsotti, Barsuglia, Barta, Bartlett, Bartos,
  Bassiri, Basti, Bawaj, Bayley, Bazzan, B\'ecsy, Bejger, Belahcene, Bell,
  Beniwal, Berger, Bergmann, Bernuzzi, Bero, Berry, Bersanetti, Bertolini,
  Betzwieser, Bhandare, Bidler, Bilenko, Bilgili, Billingsley, Birch, Birney,
  Birnholtz, Biscans, Biscoveanu, Bisht, Bitossi, Bizouard, Blackburn,
  Blackman, Blair, Blair, Blair, Bloemen, Bode, Boer, Boetzel, Bogaert, Bondu,
  Bonilla, Bonnand, Booker, Boom, Booth, Bork, Boschi, Bose, Bossie, Bossilkov,
  Bosveld, Bouffanais, Bozzi, Bradaschia, Brady, Bramley, Branchesi, Brau,
  Briant, Briggs, Brighenti, Brillet, Brinkmann, Brisson, Brockill, Brooks,
  Brown, Brunett, Buikema, Bulik, Bulten, Buonanno, Buskulic,
  Bustamante~Rosell, Buy, Byer, Cabero, Cadonati, Cagnoli, Cahillane,
  Calder\'on~Bustillo, Callister, Calloni, Camp, Campbell, Canepa, Cannon, Cao,
  Cao, Capocasa, Carbognani, Caride, Carney, Carullo, Casanueva~Diaz,
  Casentini, Caudill, Cavagli\`a, Cavalier, Cavalieri, Cella, Cerd\'a-Dur\'an,
  Cerretani, Cesarini, Chaibi, Chakravarti, Chamberlin, Chan, Chao, Charlton,
  Chase, Chassande-Mottin, Chatterjee, Chaturvedi, Chatziioannou, Cheeseboro,
  Chen, Chen, Chen, Cheng, Cheong, Chia, Chincarini, Chiummo, Cho, Cho, Cho,
  Christensen, Chu, Chua, Chung, Chung, Ciani, Ciobanu, Ciolfi, Cipriano,
  Cirone, Clara, Clark, Clearwater, Cleva, Cocchieri, Coccia, Cohadon, Cohen,
  Colgan, Colleoni, Collette, Collins, Cominsky, Constancio, Conti, Cooper,
  Corban, Corbitt, Cordero-Carri\'on, Corley, Cornish, Corsi, Cortese, Costa,
  Cotesta, Coughlin, Coughlin, Coulon, Countryman, Couvares, Covas, Cowan,
  Coward, Cowart, Coyne, Coyne, Creighton, Creighton, Cripe, Croquette,
  Crowder, Cullen, Cumming, Cunningham, Cuoco, Canton, D\'alya, Danilishin,
  D'Antonio, Danzmann, Dasgupta, Da~Silva~Costa, Datrier, Dattilo, Dave,
  Davier, Davis, Daw, DeBra, Deenadayalan, Degallaix, De~Laurentis,
  Del\'eglise, Del~Pozzo, DeMarchi, Demos, Dent, De~Pietri, Derby, De~Rosa,
  De~Rossi, DeSalvo, de~Varona, Dhurandhar, D\'{\i}az, Dietrich, Di~Fiore,
  Di~Giovanni, Di~Girolamo, Di~Lieto, Ding, Di~Pace, Di~Palma, Di~Renzo,
  Dmitriev, Doctor, Donovan, Dooley, Doravari, Dorrington, Downes, Drago,
  Driggers, Du, Ducoin, Dupej, Dwyer, Easter, Edo, Edwards, Effler, Ehrens,
  Eichholz, Eikenberry, Eisenmann, Eisenstein, Essick, Estelles, Estevez,
  Etienne, Etzel, Evans, Evans, Fafone, Fair, Fairhurst, Fan, Farinon, Farr,
  Farr, Fauchon-Jones, Favata, Fays, Fazio, Fee, Feicht, Fejer, Feng,
  Fernandez-Galiana, Ferrante, Ferreira, Ferreira, Ferrini, Fidecaro, Fiori,
  Fiorucci, Fishbach, Fisher, Fishner, Fitz-Axen, Flaminio, Fletcher, Flynn,
  Fong, Font, Forsyth, Fournier, Frasca, Frasconi, Frei, Freise, Frey, Frey,
  Fritschel, Frolov, Fulda, Fyffe, Gabbard, Gadre, Gaebel, Gair, Gammaitoni,
  Ganija, Gaonkar, Garcia, Garc\'{\i}a-Quir\'os, Garufi, Gateley, Gaudio, Gaur,
  Gayathri, Gemme, Genin, Gennai, George, George, Gergely, Germain, Ghonge,
  Ghosh, Ghosh, Ghosh, Giacomazzo, Giaime, Giardina, Giazotto, Gill, Giordano,
  Glover, Godwin, Goetz, Goetz, Goncharov, Gonz\'alez, Gonzalez~Castro,
  Gopakumar, Gorodetsky, Gossan, Gosselin, Gouaty, Grado, Graef, Granata,
  Grant, Gras, Grassia, Gray, Gray, Greco, Green, Green, Gretarsson, Groot,
  Grote, Grunewald, Gruning, Guidi, Gulati, Guo, Gupta, Gupta, Gustafson,
  Gustafson, Haegel, Halim, Hall, Hall, Hamilton, Hammond, Haney, Hanke, Hanks,
  Hanna, Hannam, Hannuksela, Hanson, Hardwick, Haris, Harms, Harry, Harry,
  Haster, Haughian, Hayes, Healy, Heidmann, Heintze, Heitmann, Hello, Hemming,
  Hendry, Heng, Hennig, Heptonstall, Hernandez~Vivanco, Heurs, Hild, Hinderer,
  Hoak, Hochheim, Hofman, Holgado, Holland, Holt, Holz, Hopkins, Horst, Hough,
  Howell, Hoy, Hreibi, Huang, Huerta, Huet, Hughey, Hulko, Husa, Huttner,
  Huynh-Dinh, Idzkowski, Iess, Ingram, Inta, Intini, Irwin, Isa, Isac, Isi,
  Iyer, Izumi, Jacqmin, Jadhav, Jani, Janthalur, Jaranowski, Jenkins, Jiang,
  Johnson, Johnson-McDaniel, Jones, Jones, Jones, Jonker, Ju, Junker,
  Kalaghatgi, Kalogera, Kamai, Kandhasamy, Kang, Kanner, Kapadia, Karki,
  Karvinen, Kashyap, Kasprzack, Katsanevas, Katsavounidis, Katzman, Kaufer,
  Kawabe, Keerthana, K\'ef\'elian, Keitel, Kennedy, Key, Khalili, Khan, Khan,
  Khan, Khan, Khazanov, Khursheed, Kijbunchoo, Kim, Kim, Kim, Kim, Kim, Kim,
  Kimball, King, King, Kinley-Hanlon, Kirchhoff, Kissel, Kleybolte, Klika,
  Klimenko, Knowles, Koch, Koehlenbeck, Koekoek, Koley, Kondrashov, Kontos,
  Koper, Korobko, Korth, Kowalska, Kozak, Kringel, Krishnendu, Kr\'olak, Kuehn,
  Kumar, Kumar, Kumar, Kumar, Kuo, Kutynia, Kwang, Lackey, Lai, Lam, Landry,
  Lane, Lang, Lange, Lantz, Lanza, Lartaux-Vollard, Lasky, Laxen, Lazzarini,
  Lazzaro, Leaci, Leavey, Lecoeuche, Lee, Lee, Lee, Lee, Lee, Lee, Lehmann,
  Lenon, Leroy, Letendre, Levin, Li, Li, Li, Li, Lin, Linde, Linker,
  Littenberg, Liu, Liu, Lo, Lockerbie, London, Longo, Lorenzini, Loriette,
  Lormand, Losurdo, Lough, Lousto, Lovelace, Lower, L\"uck, Lumaca, Lundgren,
  Lynch, Ma, Macas, Macfoy, MacInnis, Macleod, Macquet, Maga\~na Sandoval,
  Maga\~na Zertuche, Magee, Majorana, Maksimovic, Malik, Man, Mandic, Mangano,
  Mansell, Manske, Mantovani, Marchesoni, Marion, M\'arka, M\'arka, Markakis,
  Markosyan, Markowitz, Maros, Marquina, Marsat, Martelli, Martin, Martin,
  Martynov, Mason, Massera, Masserot, Massinger, Masso-Reid, Mastrogiovanni,
  Matas, Matichard, Matone, Mavalvala, Mazumder, McCann, McCarthy, McClelland,
  McCormick, McCuller, McGuire, McIver, McManus, McRae, McWilliams, Meacher,
  Meadors, Mehmet, Mehta, Meidam, Melatos, Mendell, Mercer, Mereni, Merilh,
  Merzougui, Meshkov, Messenger, Messick, Metzdorff, Meyers, Miao, Michel,
  Middleton, Mikhailov, Milano, Miller, Miller, Millhouse, Mills,
  Milovich-Goff, Minazzoli, Minenkov, Mishkin, Mishra, Mistry, Mitra,
  Mitrofanov, Mitselmakher, Mittleman, Mo, Moffa, Mogushi, Mohapatra, Montani,
  Moore, Moraru, Moreno, Morisaki, Mours, Mow-Lowry, Mukherjee, Mukherjee,
  Mukherjee, Mukund, Mullavey, Munch, Mu\~niz, Muratore, Murray, Nagar,
  Nardecchia, Naticchioni, Nayak, Neilson, Nelemans, Nelson, Nery, Neunzert,
  Ng, Ng, Nguyen, Nichols, Nielsen, Nissanke, Nitz, Nocera, North, Nuttall,
  Obergaulinger, Oberling, O'Brien, O'Dea, Ogin, Oh, Oh, Ohme, Ohta, Okada,
  Oliver, Oppermann, Oram, O'Reilly, Ormiston, Ortega, O'Shaughnessy, Ossokine,
  Ottaway, Overmier, Owen, Pace, Pagano, Page, Pai, Pai, Palamos, Palashov,
  Palomba, Pal-Singh, Pan, Pang, Pang, Pankow, Pannarale, Pant, Paoletti,
  Paoli, Papa, Parida, Parker, Pascucci, Pasqualetti, Passaquieti, Passuello,
  Patil, Patricelli, Pearlstone, Pedersen, Pedraza, Pedurand, Pele, Penn,
  Perego, Perez, Perreca, Pfeiffer, Phelps, Phukon, Piccinni, Pichot,
  Piergiovanni, Pillant, Pinard, Pirello, Pitkin, Poggiani, Pong, Ponrathnam,
  Popolizio, Porter, Powell, Prajapati, Prasad, Prasai, Prasanna, Pratten,
  Prestegard, Privitera, Prodi, Prokhorov, Puncken, Punturo, Puppo, P\"urrer,
  Qi, Quetschke, Quinonez, Quintero, Quitzow-James, Raab, Radkins, Radulescu,
  Raffai, Raja, Rajan, Rajbhandari, Rakhmanov, Ramirez, Ramos-Buades, Rana,
  Rao, Rapagnani, Raymond, Razzano, Read, Regimbau, Rei, Reid, Reitze, Ren,
  Ricci, Richardson, Richardson, Ricker, Riemenschneider, Riles, Rizzo,
  Robertson, Robie, Robinet, Rocchi, Rolland, Rollins, Roma, Romanelli, Romano,
  Romel, Romie, Rose, Rosi\ifmmode~\acute{n}\else \'{n}\fi{}ska, Rosofsky,
  Ross, Rowan, R\"udiger, Ruggi, Rutins, Ryan, Sachdev, Sadecki, Sakellariadou,
  Salafia, Salconi, Saleem, Salemi, Samajdar, Sammut, Sanchez, Sanchez,
  Sanchis-Gual, Sandberg, Sanders, Santiago, Sarin, Sassolas, Sathyaprakash,
  Saulson, Sauter, Savage, Schale, Scheel, Scheuer, Schmidt, Schnabel,
  Schofield, Sch\"onbeck, Schreiber, Schulte, Schutz, Schwalbe, Scott, Scott,
  Seidel, Sellers, Sengupta, Sennett, Sentenac, Sequino, Sergeev, Setyawati,
  Shaddock, Shaffer, Shahriar, Shaner, Shao, Sharma, Shawhan, Shen, Shink,
  Shoemaker, Shoemaker, ShyamSundar, Siellez, Sieniawska, Sigg, Silva, Singer,
  Singh, Singhal, Sintes, Sitmukhambetov, Skliris, Slagmolen, Slaven-Blair,
  Smith, Smith, Somala, Son, Sorazu, Sorrentino, Souradeep, Sowell, Spencer,
  Srivastava, Srivastava, Staats, Stachie, Standke, Steer, Steinke,
  Steinlechner, Steinlechner, Steinmeyer, Stevenson, Stocks, Stone, Stops,
  Strain, Stratta, Strigin, Strunk, Sturani, Stuver, Sudhir, Summerscales, Sun,
  Sunil, Suresh, Sutton, Swinkels, Szczepa\ifmmode~\acute{n}\else
  \'{n}\fi{}czyk, Tacca, Tait, Talbot, Talukder, Tanner, T\'apai, Taracchini,
  Tasson, Taylor, Thies, Thomas, Thomas, Thondapu, Thorne, Thrane, Tiwari,
  Tiwari, Tiwari, Toland, Tonelli, Tornasi, Torres-Forn\'e, Torrie, T\"oyr\"a,
  Travasso, Traylor, Tringali, Trovato, Trozzo, Trudeau, Tsang, Tse, Tso,
  Tsukada, Tsuna, Tuyenbayev, Ueno, Ugolini, Unnikrishnan, Urban, Usman,
  Vahlbruch, Vajente, Valdes, van Bakel, van Beuzekom, van~den Brand, Van
  Den~Broeck, Vander-Hyde, van Heijningen, van~der Schaaf, van Veggel, Vardaro,
  Varma, Vass, Vas\'uth, Vecchio, Vedovato, Veitch, Veitch, Venkateswara,
  Venugopalan, Verkindt, Vetrano, Vicer\'e, Viets, Vine, Vinet, Vitale, Vo,
  Vocca, Vorvick, Vyatchanin, Wade, Wade, Wade, Walet, Walker, Wallace, Walsh,
  Wang, Wang, Wang, Wang, Wang, Ward, Warden, Warner, Was, Watchi, Weaver, Wei,
  Weinert, Weinstein, Weiss, Wellmann, Wen, Wessel, We\ss{}els, Westhouse,
  Wette, Whelan, White, Whiting, Whittle, Wilken, Williams, Williamson, Willis,
  Willke, Wimmer, Winkler, Wipf, Wittel, Woan, Woehler, Wofford, Worden,
  Wright, Wu, Wysocki, Xiao, Yamamoto, Yancey, Yang, Yap, Yazback, Yeeles, Yu,
  Yu, Yuen, Yvert, Zadro\ifmmode~\dot{z}\else \.{z}\fi{}ny, Zanolin, Zappa,
  Zelenova, Zendri, Zevin, Zhang, Zhang, Zhang, Zhao, Zhou, Zhou, Zhu,
  Zimmerman, Zlochower, Zucker, \& Zweizig}]{abb19}
Abbott, B.~P., Abbott, R., Abbott, T.~D., {et~al.} 2019, Phys. Rev. X, 9,
  031040

\bibitem[{{Bavera} {et~al.}(2020){Bavera}, {Fragos}, {Qin}, {Zapartas},
  {Neijssel}, {Mandel}, {Batta}, {Gaebel}, {Kimball}, \& {Stevenson}}]{bav20}
{Bavera}, S.~S., {Fragos}, T., {Qin}, Y., {et~al.} 2020, \aap, 635, A97

\bibitem[{{Bavera} {et~al.}(2023){Bavera}, {Fragos}, {Zapartas}, {Andrews},
  {Kalogera}, {Berry}, {Kruckow}, {Dotter}, {Kovlakas}, {Misra}, {Rocha},
  {Srivastava}, {Sun}, \& {Xing}}]{bav23}
{Bavera}, S.~S., {Fragos}, T., {Zapartas}, E., {et~al.} 2023, Nature Astronomy,
  7, 1090

\bibitem[{{Belczynski} {et~al.}(2008){Belczynski}, {Kalogera}, {Rasio}, {Taam},
  {Zezas}, {Bulik}, {Maccarone}, \& {Ivanova}}]{bel08}
{Belczynski}, K., {Kalogera}, V., {Rasio}, F.~A., {et~al.} 2008, \apjs, 174,
  223

\bibitem[{{Belczynski} {et~al.}(2020){Belczynski}, {Klencki}, {Fields},
  {Olejak}, {Berti}, {Meynet}, {Fryer}, {Holz}, {O'Shaughnessy}, {Brown},
  {Bulik}, {Leung}, {Nomoto}, {Madau}, {Hirschi}, {Kaiser}, {Jones}, {Mondal},
  {Chruslinska}, {Drozda}, {Gerosa}, {Doctor}, {Giersz}, {Ekstrom}, {Georgy},
  {Askar}, {Baibhav}, {Wysocki}, {Natan}, {Farr}, {Wiktorowicz}, {Coleman
  Miller}, {Farr}, \& {Lasota}}]{bel20}
{Belczynski}, K., {Klencki}, J., {Fields}, C.~E., {et~al.} 2020, \aap, 636,
  A104

\bibitem[{{Burkart} {et~al.}(2012){Burkart}, {Quataert}, {Arras}, \&
  {Weinberg}}]{bur12}
{Burkart}, J., {Quataert}, E., {Arras}, P., \& {Weinberg}, N.~N. 2012, \mnras,
  421, 983

\bibitem[{{Claret}(2004)}]{cla04}
{Claret}, A. 2004, \aap, 424, 919

\bibitem[{{Eggenberger} {et~al.}(2008){Eggenberger}, {Meynet}, {Maeder},
  {Hirschi}, {Charbonnel}, {Talon}, \& {Ekstr{\"o}m}}]{egg08}
{Eggenberger}, P., {Meynet}, G., {Maeder}, A., {et~al.} 2008, \apss, 316, 43

\bibitem[{{Ekstr{\"o}m} {et~al.}(2012){Ekstr{\"o}m}, {Georgy}, {Eggenberger},
  {Meynet}, {Mowlavi}, {Wyttenbach}, {Granada}, {Decressin}, {Hirschi},
  {Frischknecht}, {Charbonnel}, \& {Maeder}}]{eks12}
{Ekstr{\"o}m}, S., {Georgy}, C., {Eggenberger}, P., {et~al.} 2012, \aap, 537,
  A146

\bibitem[{{Fragos} {et~al.}(2023){Fragos}, {Andrews}, {Bavera}, {Berry},
  {Coughlin}, {Dotter}, {Giri}, {Kalogera}, {Katsaggelos}, {Kovlakas},
  {Lalvani}, {Misra}, {Srivastava}, {Qin}, {Rocha}, {Rom{\'a}n-Garza}, {Serra},
  {Stahle}, {Sun}, {Teng}, {Trajcevski}, {Tran}, {Xing}, {Zapartas}, \&
  {Zevin}}]{fra23}
{Fragos}, T., {Andrews}, J.~J., {Bavera}, S.~S., {et~al.} 2023, \apjs, 264, 45

\bibitem[{{Fragos} \& {McClintock}(2015)}]{fra15}
{Fragos}, T. \& {McClintock}, J.~E. 2015, \apj, 800, 17

\bibitem[{{Fuller}(2017)}]{ful17}
{Fuller}, J. 2017, \mnras, 472, 1538

\bibitem[{{Hotokezaka} \& {Piran}(2017)}]{hot17}
{Hotokezaka}, K. \& {Piran}, T. 2017, \apj, 842, 111

\bibitem[{{Hurley} {et~al.}(2002){Hurley}, {Tout}, \& {Pols}}]{hur02}
{Hurley}, J.~R., {Tout}, C.~A., \& {Pols}, O.~R. 2002, \mnras, 329, 897

\bibitem[{{Hut}(1981)}]{hut81}
{Hut}, P. 1981, \aap, 99, 126

\bibitem[{{Kushnir} {et~al.}(2016){Kushnir}, {Zaldarriaga}, {Kollmeier}, \&
  {Waldman}}]{kus16}
{Kushnir}, D., {Zaldarriaga}, M., {Kollmeier}, J.~A., \& {Waldman}, R. 2016,
  \mnras, 462, 844

\bibitem[{{Ma} \& {Fuller}(2021)}]{ma21}
{Ma}, L. \& {Fuller}, J. 2021, \apj, 918, 16

\bibitem[{{Maeder} \& {Meynet}(2004)}]{mae04}
{Maeder}, A. \& {Meynet}, G. 2004, \aap, 422, 225

\bibitem[{{Mirouh} {et~al.}(2023){Mirouh}, {Hendriks}, {Dykes}, {Moe}, \&
  {Izzard}}]{mir23}
{Mirouh}, G.~M., {Hendriks}, D.~D., {Dykes}, S., {Moe}, M., \& {Izzard}, R.~G.
  2023, \mnras, 524, 3978

\bibitem[{{Packet}(1981)}]{pac81}
{Packet}, W. 1981, \aap, 102, 17

\bibitem[{{Paxton} {et~al.}(2015){Paxton}, {Marchant}, {Schwab}, {Bauer},
  {Bildsten}, {Cantiello}, {Dessart}, {Farmer}, {Hu}, {Langer}, {Townsend},
  {Townsley}, \& {Timmes}}]{pax15}
{Paxton}, B., {Marchant}, P., {Schwab}, J., {et~al.} 2015, \apjs, 220, 15

\bibitem[{{Qin} {et~al.}(2018){Qin}, {Fragos}, {Meynet}, {Andrews},
  {S{\o}rensen}, \& {Song}}]{qin18}
{Qin}, Y., {Fragos}, T., {Meynet}, G., {et~al.} 2018, \aap, 616, A28

\bibitem[{{Sepinsky} {et~al.}(2007){Sepinsky}, {Willems}, {Kalogera}, \&
  {Rasio}}]{sep07}
{Sepinsky}, J.~F., {Willems}, B., {Kalogera}, V., \& {Rasio}, F.~A. 2007, \apj,
  667, 1170

\bibitem[{{Siess} {et~al.}(2013){Siess}, {Izzard}, {Davis}, \&
  {Deschamps}}]{sie13}
{Siess}, L., {Izzard}, R.~G., {Davis}, P.~J., \& {Deschamps}, R. 2013, \aap,
  550, A100

\bibitem[{{Song} {et~al.}(2013){Song}, {Maeder}, {Meynet}, {Huang},
  {Ekstr{\"o}m}, \& {Granada}}]{son13}
{Song}, H.~F., {Maeder}, A., {Meynet}, G., {et~al.} 2013, \aap, 556, A100

\bibitem[{{Song} {et~al.}(2016){Song}, {Meynet}, {Maeder}, {Ekstr{\"o}m}, \&
  {Eggenberger}}]{son16}
{Song}, H.~F., {Meynet}, G., {Maeder}, A., {Ekstr{\"o}m}, S., \& {Eggenberger},
  P. 2016, \aap, 585, A120

\bibitem[{{Song} {et~al.}(2018){Song}, {Meynet}, {Maeder}, {Ekstr{\"o}m},
  {Eggenberger}, {Georgy}, {Qin}, {Fragos}, {Soerensen}, {Barblan}, \&
  {Wade}}]{son18}
{Song}, H.~F., {Meynet}, G., {Maeder}, A., {et~al.} 2018, \aap, 609, A3

\bibitem[{{Spruit}(2002)}]{spr02}
{Spruit}, H.~C. 2002, \aap, 381, 923

\bibitem[{Su \& Lai(2021)}]{su21}
Su, Y. \& Lai, D. 2021, Monthly Notices of the Royal Astronomical Society, 510,
  4943

\bibitem[{{Toonen} {et~al.}(2016){Toonen}, {Hamers}, \& {Portegies
  Zwart}}]{too16}
{Toonen}, S., {Hamers}, A., \& {Portegies Zwart}, S. 2016, Computational
  Astrophysics and Cosmology, 3, 6

\bibitem[{{Witte} \& {Savonije}(2002)}]{wit02}
{Witte}, M.~G. \& {Savonije}, G.~J. 2002, \aap, 386, 222

\bibitem[{{Zahn}(1977)}]{zah77}
{Zahn}, J.~P. 1977, \aap, 57, 383

\bibitem[{{Zaldarriaga} {et~al.}(2018){Zaldarriaga}, {Kushnir}, \&
  {Kollmeier}}]{zal18}
{Zaldarriaga}, M., {Kushnir}, D., \& {Kollmeier}, J.~A. 2018, \mnras, 473, 4174

\bibitem[{{Zanazzi} \& {Wu}(2021)}]{zan21}
{Zanazzi}, J.~J. \& {Wu}, Y. 2021, \aj, 161, 263

\end{thebibliography}
\section{Comparison of $\tau_{{\rm sync},s_{22}}$ and $\tau_{{\rm sync,hur}}$}\label{AppA}
\begin{figure}[h]
\centering
\centerline{\includegraphics[trim=0cm 0cm 2cm 1.5cm, clip=true, width=1.02\columnwidth,angle=0]{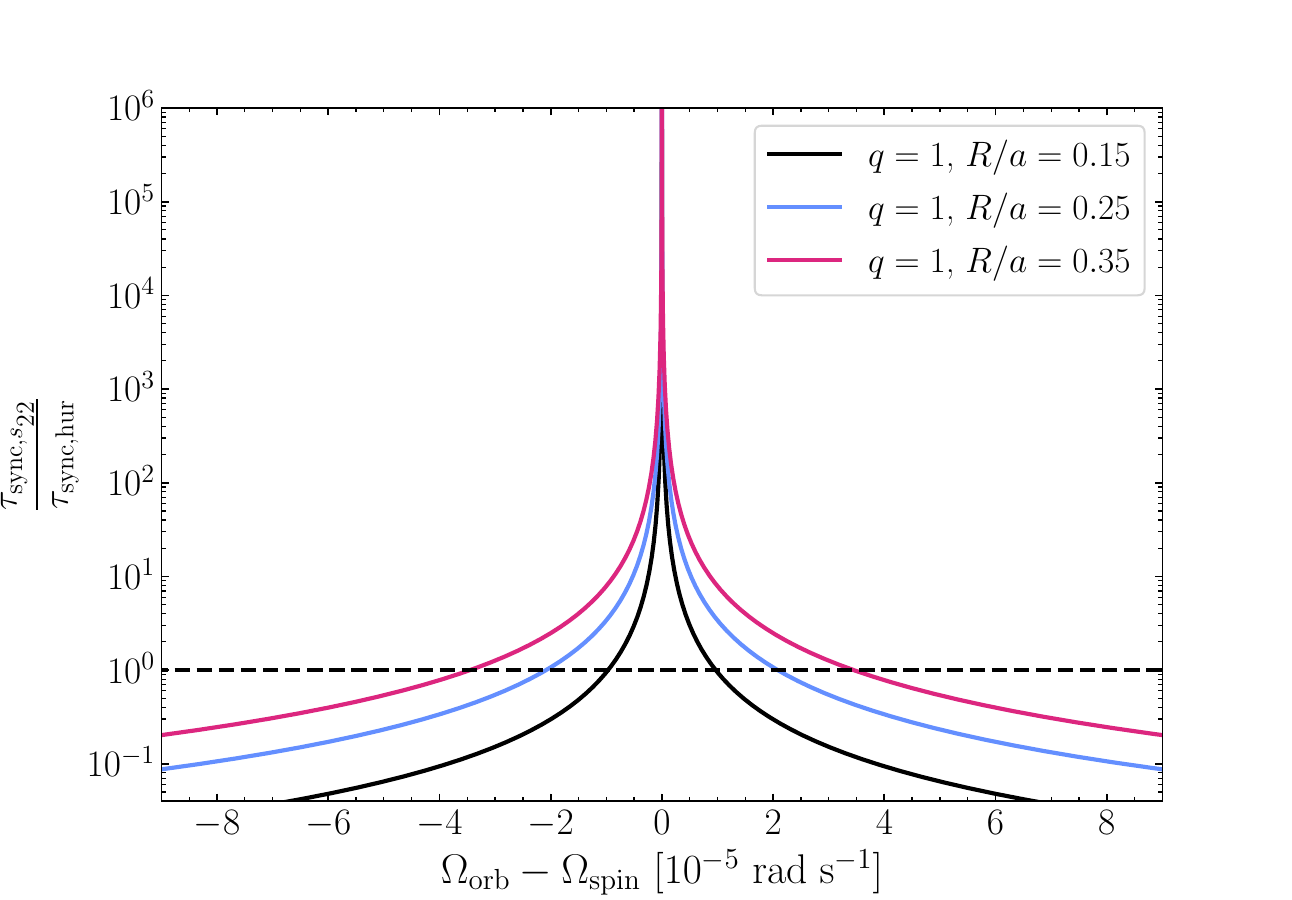}}
\caption{$\frac{\tau_{{\rm sync},s_{22}}}{\tau_{{\rm sync,hur}}}$ as a function of the difference $\Omega_{\rm orb}-\Omega_{\rm spin}$ with $q=1$ and a few values of $R/a$ (solid lines). The solid lines cross the horizontal dashed line when the  two timescales are equal.
}
\label{tausyncs}
\end{figure}

\begin{figure}[h]
\centering
\centerline{\includegraphics[trim=0cm 0cm 2cm 1.5cm, clip=true, width=1.02\columnwidth,angle=0]{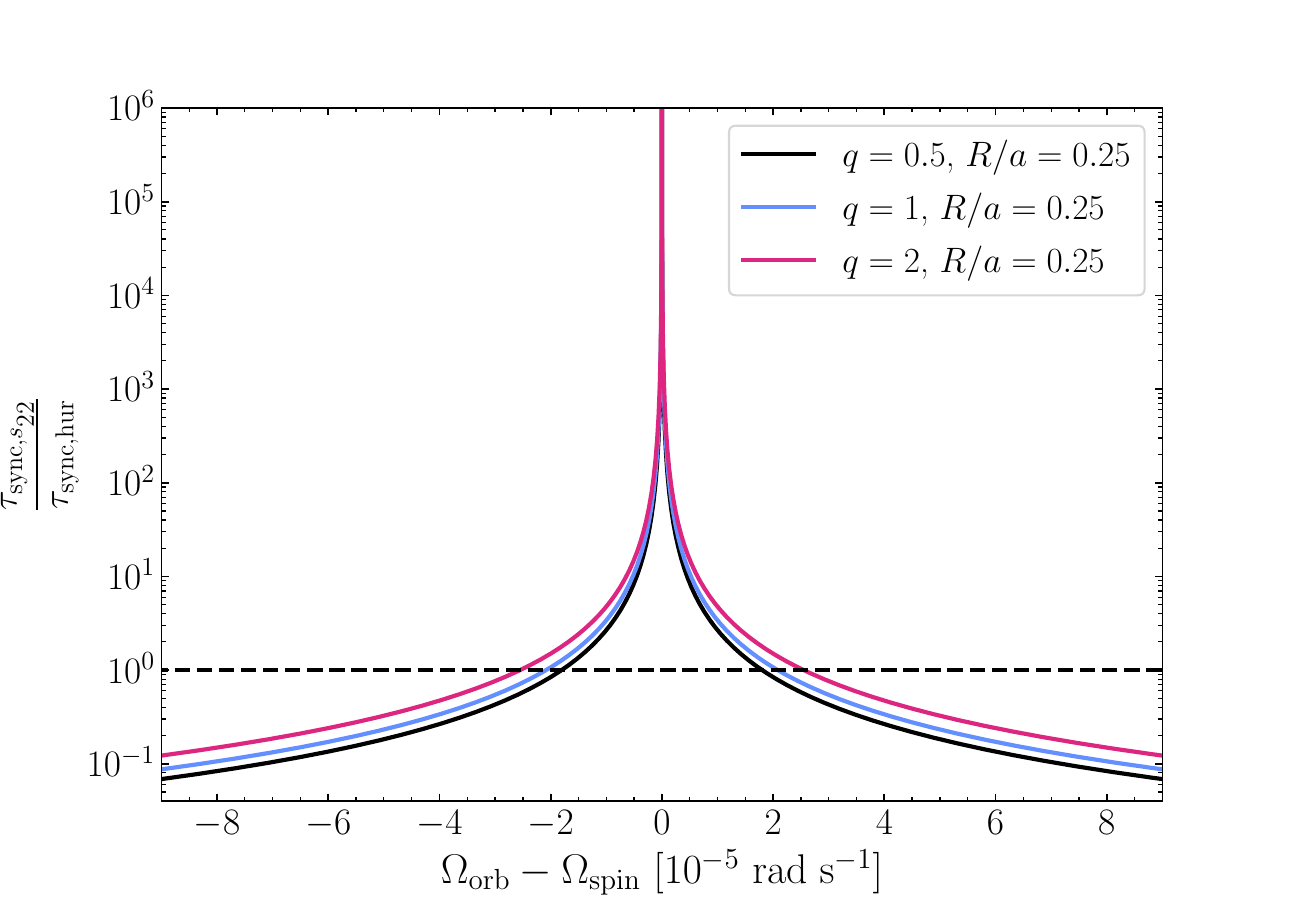}}
\caption{
$\frac{\tau_{{\rm sync},s_{22}}}{\tau_{{\rm sync,hur}}}$ as a function of the difference $\Omega_{\rm orb}-\Omega_{\rm spin}$ with $R/a=0.25$ and a few values of $q$ (solid lines). The solid lines cross the horizontal dashed line when the   two timescales are equal.}
\label{tausyncs_q}
\end{figure}

Looking at the expressions of $\tau_{{\rm sync},s_{22}}$ and $\tau_{{\rm sync,hur}}$, we find that $\tau_{{\rm sync},s_{22}}$ is dependent on $s_{22}$ (i.e., on the difference in $\Omega$), which is not the case for $\tau_{{\rm sync,hur}}$. The ratio of $\tau_{{\rm sync},s_{22}}$ to $\tau_{{\rm sync,hur}}$ can be obtained:
\begin{equation}
    \frac{\tau_{{\rm sync},s_{22}}}{\tau_{{\rm sync,hur}}}=(1+q)^{5/6}\left(\frac{R}{a}\right)^{5/2}s_{22}^{-5/3}\text{sgn}(s_{22}).
\end{equation}
We note that $\tau_{{\rm sync},s_{22}}$ has been corrected with a sign function in order to be positive no matter the sign of $s_{22}$ (see comments in Appendix \ref{AppB}).
The terms in $q$ and $R/a$ come from the Hurley calculation, the term in $s_{22}$ from the original timescale. Since only the original timescale depends on $s_{22}$, the ratio of the two timescales has a strong dependence on the difference in $\Omega$. When the star is far from synchronization, $\tau_{{\rm sync},s_{22}}$ is shorter. When the star is close to synchronization, $\tau_{{\rm sync},s_{22}}$ diverges. The ratio $\frac{\tau_{{\rm sync},s_{22}}}{\tau_{{\rm sync,hur}}}$ is shown as a function of the difference in $\Omega$ with $q=1$ and a few values of $R/a$ in Fig. \ref{tausyncs} and with $R/a=0.25$ and a few values of $q$ in Fig. \ref{tausyncs_q}.
Figures \ref{tausyncs} and \ref{tausyncs_q} show that, depending on the value of the difference in $\Omega$, the values of the synchronization timescales $\tau_{{\rm sync},s_{22}}$ and ${\tau_{{\rm sync,hur}}}$ can differ by several orders of magnitude. When $\Omega_{\rm spin}\to\Omega_{\rm orb}$, the ratio of the two timescales diverges. The present plots also show that only for a single value of the difference in $\Omega$ are the two timescales equal. This is all very consistent with the results shown in the analytical part (section \ref{analytical}).
\section{Derivation of Eq. \eqref{zahn_corrected}}\label{AppB}
The first issue with the angular velocity equation in \eqref{zahn} is that it never provides a negative torque, as the power 8/3 in $s_{22}$ implies that the expression $s_{22}^{8/3}$ is positive no matter the sign of the difference between $\Omega_{\rm spin}$ and $\Omega_{\rm orb}$. In order for the torque to always act toward synchronization, Eq. \eqref{zahn} should be rewritten as
\begin{equation}
\restriction{\frac{\text{d}}{\text{d}t}\left(I\Omega_{\rm spin}\right)}{\rm Dyn}=\frac{3}{2}\frac{GM^2}{R}E_2\left(q^2\left(\frac{R}{a}\right)^6\right)s_{22}^{8/3}\text{sgn}(s_{22}),
\label{correct}
\end{equation}
as has been done by  \citet{su21}, among others. \par The second issue is that the angular velocity equation (equation (5.6) in \Zah\ ) neglects the corrections called for when the orbit is eccentric. In order to obtain an equation valid for nonzero eccentricities, we use equation (3.8) in \Zah\, with $$\varepsilon_n^{lm}=E_ns_{lm}^{8/3}\text{sgn}(s_{lm}),$$ which corresponds to equation (5.5) in their paper corrected with a sign function for each term. Applying this definition for $\varepsilon_n^{lm}$ to the secular equations (3.7) and (3.8) of \Zah\ provides Eq. \eqref{zahn_corrected}.
\section{Spin-up case}\label{AppC}
\begin{figure}[h]
\centering
\centerline{\includegraphics[trim=2cm 1cm 2cm 1.5cm, clip=true, width=1.02\columnwidth,angle=0]{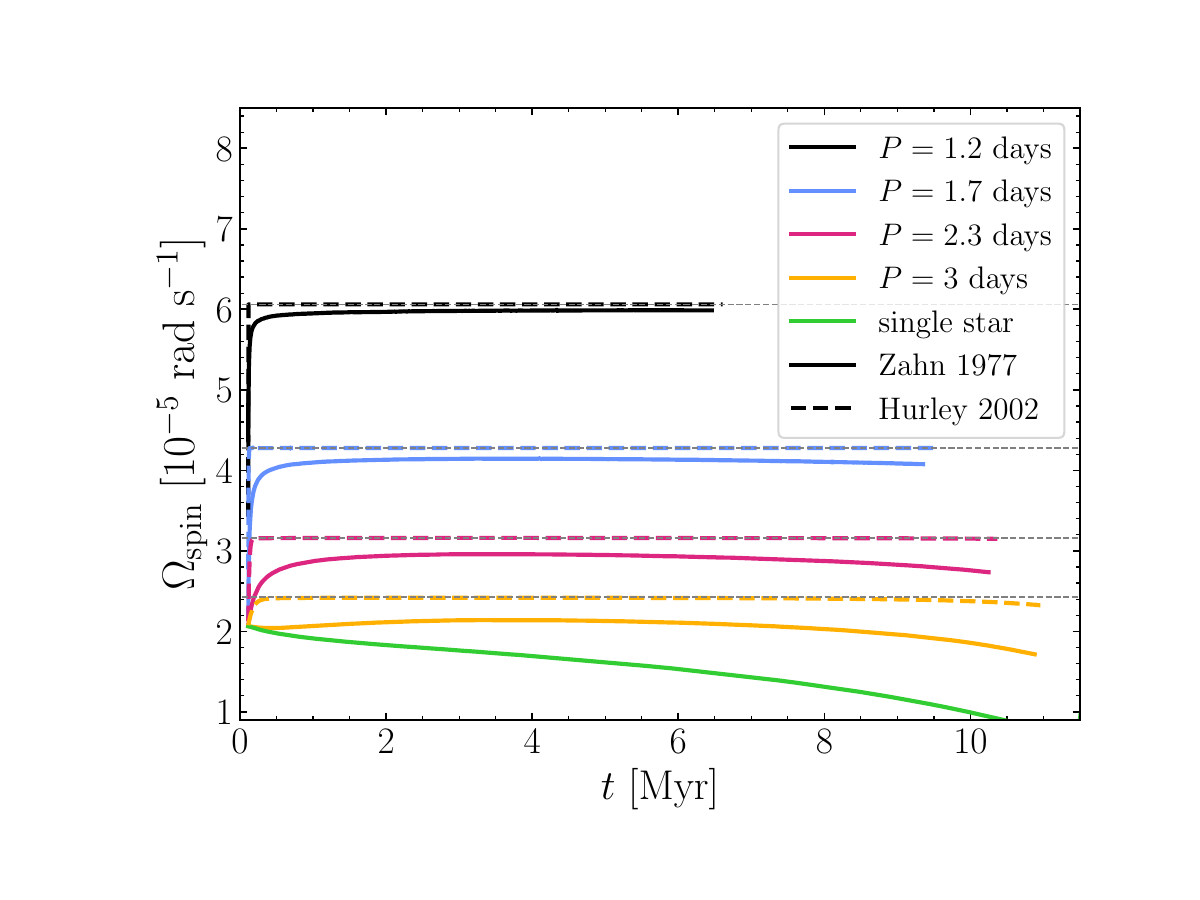}}
\caption{
MS $\Omega_{\rm spin}$ evolution in the spin-up case for stellar models with initial velocity $v/v_{\rm crit}=0.1$ and orbital periods $P=[1.2,1.7,2.3,3]$ days with the tidal torques defined in Eqs. \eqref{zahn_torque} (Zahn 1977, solid line) and \eqref{hurley_torque} (Hurley 2002, dashed line). The grey dashed lines represent the orbital angular velocities corresponding to the selected periods. The single star evolution is shown as a reference.}
\label{spin_up}
\end{figure}

In this section we present the evolution of the surface angular velocity for the same stellar and binarity parameters as in the previous section, but with $v_0/v_{\rm crit}=0.1$ as initial velocity for the stellar models (i.e., an angular velocity of $\Omega_0=2.1\cdot 10^{-5}$ rad/s). We compare the evolution of the surface angular velocity depending on the   formula used for the torque, and we show as a reference the evolution of the surface angular velocity for the single star evolution. The time evolutions of $\Omega_{\rm spin}$ are shown in Fig. \ref{spin_up}.
\noindent As in the spin-down case, we note that the Zahn torque is less efficient in bringing the star to synchronization. The difference is more pronounced in this case as the initial spin of the star is smaller than the orbital spin. The torque must act against the natural evolution of $\Omega_{\rm spin}$, whereas in the spin-down case they are (initially) acting in the same direction. In the spin-up case we observe that the rotational spins actually never reach the orbital spins with the Zahn torque (even though pseudo-synchronization is visible for the smallest periods). 
\end{appendix}

\end{document}